\documentclass[a4paper,10pt,twoside]{cpc-hepnp}
\usepackage{multicol}
\usepackage{graphicx}
\usepackage{booktabs}
\usepackage{amssymb,bm,mathrsfs,bbm,amscd}
\usepackage[tbtags]{amsmath}
\usepackage{lastpage}
\usepackage{CJK}
\usepackage{color}
\usepackage{epstopdf}
\usepackage{epsfig}
\usepackage{ulem}
\usepackage{pdfsync}

\begin{document}
\begin{CJK*}{GB}{gbsn}
\title{Three-quark potential at finite temperature and chemical potential}
\author{\quad Jia-Jie Jiang(½­¼Ñ½Ü)$^{1}$
\quad Ya-Zhao Xiao(ФÑdz¯)$^{1}$
\quad Jiajia Qin(ÇؼѼÑ)$^{1;3)}$
\quad Xiaohua Li(ÀîС»ª)$^{1;2)}$
\quad Xun Chen(³ÂÑ«)$^{1;1)}$\\
\email{chenxunhep@qq.com;corresponding author}
\email{lixiaohuaphysics@126.com}
\email{jiajiaqin@usc.edu.cn}
}
\maketitle

\address{$^1$ School of Nuclear Science and Technology, University of South China, Hengyang 421001, China\\}

\begin{abstract}
Using gauge/gravity duality, we study the potential energy and the melting of triply heavy baryon at finite temperature and chemical potential in this paper. First, we calculate three-quark potential and compare the results with quark-antiquark potential. With the increase of temperature and chemical potential, the potential energy will decrease at large distances. It is found that the three-quark potential will have an endpoint at high temperature and/or large chemical potential, which means triply heavy baryons will melt at enough high temperature and/or large chemical potential. We also discuss screening distance which can be extracted from the three-quark potential. At last, we draw the melting diagram of triply heavy baryons in the $T-\mu$ plane.
\end{abstract}

\begin{multicols}{2}

\section{Introduction}\label{int}
Triply heavy baryons are systems of great theoretical interest, since they may help us to better understand the interaction among heavy quarks in an environment free of valence light quarks\cite{Flynn:2011gf}. The challenge is to explain the structure and properties of such baryons in the theoretical side. It is expected that the potential models would be useful to gain important insights into understanding how baryons are put together from quarks. The three-quark potential is one of the most important inputs of the potential models and also a key to understand the quark confinement mechanism in baryons\cite{Andreev:2015riv}.

The interest in studying the interquark potential for a three-quark system has a long history due to its importance in the
spectroscopy of baryons. Besides, it is also interesting to investigate how the three-quark potential is different from or the same as the two-body quark-antiquark potential. The first studies date back to the mid '80s\cite{Sommer:1984xq,Sommer:1985da} and
after more than a decade a new turn of research started around the year 2000
which continues till now\cite{Bali:2000gf,Brambilla:2009cd,Eichmann:2016yit,Borisenko:2018zzd,Richard:2012xw}. Especially, three-quark potential has been widely investigated in the lattice QCD\cite{deForcrand:2005vv,Bornyakov:2004uv,Sakumichi:2015rfa,Bernard:2000gd,Takahashi:2000te,Matsufuru:2000qm,Takahashi:2001nw,Alexandrou:2001ip,Takahashi:2002bw,Takahashi:2003ty,Takahashi:2004rw,Suganuma:2015xby,Alexandrou:2002sn,Koma:2017hcm}. In SU(3) quenched lattice QCD, 3-quark potentials are found to be well reproduced by
\begin{equation}\label{QQQformula}
V_{3 \mathrm{Q}}\left(\mathbf{r}_{1}, \mathbf{r}_{2}, \mathbf{r}_{3}\right)=\sigma_{3 \mathrm{Q}} L_{\min }-\sum_{i<j} \frac{A_{3 \mathrm{Q}}}{\left|\mathbf{r}_{i}-\mathbf{r}_{j}\right|}+C_{3 \mathrm{Q}}.
\end{equation}
Here, $\mathbf{r}_{1}, \mathbf{r}_{2}$, and $\mathbf{r}_{3}$ are the positions of the three quarks, and $L_{\min }$ is the minimum flux-tube length connecting the three quarks. The strength of quark confinement is controlled by the string tension $\sigma_{3 \mathrm{Q}}$ of the flux tube. The form of Eq.~(\ref{QQQformula}) is called the Y ansatz. These functional forms~(\ref{QQQformula}) indicate the flux-tube picture on the confinement mechanism. Valence quarks are linked by the color flux tube as a quasi-one-dimensional object\cite{Sakumichi:2015rfa}.

The difficulty in deriving quark confinement directly from QCD is due to the non-perturbative features of QCD. With the discovery of AdS/CFT correspondence, using the classical gravitational theory to solve the non-perturbative QCD problems becomes possible\cite{Maldacena:1997re,Gubser:1998bc,Witten:1998qj}. The holographic quark-antiquark potential has been first studied in Ref.~\cite{Maldacena:1998im}. Then, Ref.~\cite{Rey:1998bq} extended the quark-antiquark potential at finite temperature. In Refs.~\cite{Andreev:2006nw,He:2010ye,Colangelo:2010pe,Li:2011hp}, they introduce a deformed factor to reproduce the Cornell potential of the quark-antiquark pair. After that, the quark-antiquark potential in the extreme conditions has been investigated in Refs.~\cite{Fadafan:2012qy,Chakraborty:2012dt,Finazzo:2014rca,Zhang:2015faa,Gursoy:2020kjd,Chen:2017lsf,Zhou:2020ssi,Zhou:2021sdy}.

Recently, effective multi-quark potential models from holography have been proposed by Oleg Andreev at vanishing temperature and chemical potential\cite{Andreev:2015riv,Andreev:2015iaa,Andreev:2019cbc,Andreev:2020xor,Andreev:2021bfg,Andreev:2021eyj,Andreev:2022cax}. As we know, the environment of high temperature and density is formed in relativistic heavy-ion collision experiments. Based on the above discussion, we recently studied the doubly heavy baryon at finite temperature from holography\cite{Chen:2021bkc}. In this paper, we continue the last work and investigate the potential energy of the triply heavy baryon at finite temperature and chemical potential.

The rest parts of this paper are organized as follows. In Sec. \textmd{\ref{Setup}}, we review the quark-antiquark potential and introduce the calculation of the three-quark potential. In Sec. \textmd{\ref{Numerical results}}, we present the numerical results and compare the quark-antiquark potential with three-quark potential. The summary and conclusion are given in Sec. \textmd{\ref{Summary and conclusion}}.
\section{Setup}\label{Setup}
In this section, we first briefly review the calculation of quark-antiquark potential at finite temperature and chemical potential. Then, we extend the calculation to the case of the three-quark potential. The possible configurations of the quark-antiquark pair and triply heavy baryon are shown in Fig.\ref{compare2and3}. Since the ground state is dominated by the most symmetric string configuration[43], we will focus on the symmetric configurations in this paper. The study of various asymmetric configurations for QQQ is an interesting topic, we may leave it for future work.
\end{multicols}
\begin{center}
	\includegraphics[width=16cm]{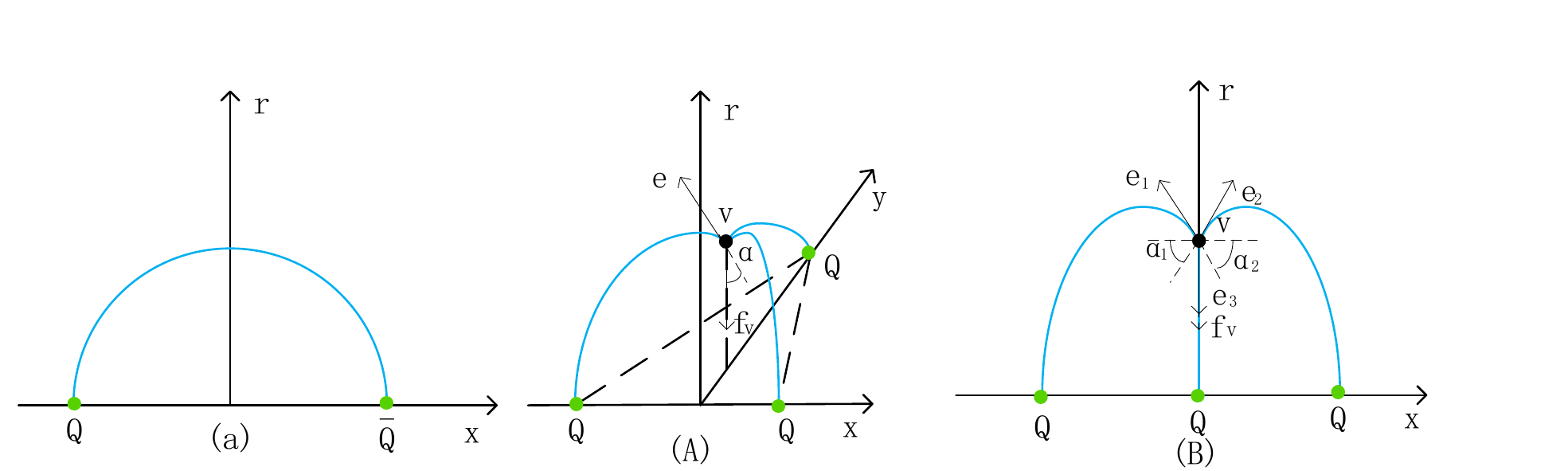}
    \figcaption{\label{compare2and3} String configurations for the $\rm Q\bar{Q}$ and the QQQ system. The heavy quarks are at the vertex of the equilateral triangle for QQQ. (A) and (B) are different configurations for QQQ.}
\end{center}
\begin{multicols}{2}
\subsection{Quark-antiquark potential}
Despite the deformed AdS-RN metric is not a self-consistent solution of Einstein equation, we can still gain some insights into problems for which there are no predictions from phenomenology and the lattice. Besides, the results provided by this model on the quark-antiquark and three-quark potentials are consistent with the lattice calculations at vanishing temperature\cite{Andreev:2015riv}. Moreover, analytic formulas can be obtained at vanishing temperature in this model. The deformed AdS-RN metric\cite{Andreev:2015riv,Andreev:2006nw} is given by
\begin{equation}\label{metric0}
\begin{gathered}
d s^{2}=e^{s r^{2}} \frac{R^{2}}{r^{2}}\left[f(r) d t^{2}+d \vec{x}^{2}+f^{-1}(r) d r^{2}\right] \\
f(r)=1-\left(\frac{1}{r_{h}^{4}}+q^{2} r_{h}^{2}\right) r^{4}+q^{2} r^{6}.
\end{gathered}
\end{equation}
Here, $q$ is the black hole charge, $r_{h}$ is the position of the black hole horizon. The Hawking temperature of the black hole is defined as
\begin{equation}
T=\frac{1}{4 \pi}\left|\frac{d f}{d r}\right|_{r=r_{h}}=\frac{1}{\pi r_{h}}\left(1-\frac{1}{2} Q^{2}\right),
\end{equation}
where $Q=q r_{h}^{3}$ and $0 \leq Q \leq \sqrt{2}$. The relation between the parameters in the deformed AdS-RN metric and the chemical potential can be obtained. Observing that, on dimensional grounds, the low $r$ behavior of the bulk gauge field $A_0(r)$ is $A_0(r) = \mu - \eta r^2$ with $\eta = \kappa q$ and $\kappa$ is a dimensionless parameter. Together with the condition that $A_0$ vanishes at the horizon $A_0(r_h) = 0$  and $A_{0}(0)=\mu$, we can determine a relation between $\mu$ and the black-hole charge $q$:
\begin{equation}
\mu= q \kappa r_h^2 = \kappa \frac{Q}{r_{h}},
\end{equation}
and we fix the parameter $\kappa$ to 1 in this paper. According to the AdS/CFT dictionary, $\mu$ is the baryon number chemical potential. Thus, we can get
\begin{equation}
\begin{gathered}
f(r)=1-\left(\frac{1}{r_{h}^{4}}+\frac{\mu^{2}}{r_{h}^{2}}\right) r^{4}+\frac{\mu^{2}}{r_{h}^{4}} r^{6}, \\
T=\frac{1}{\pi r_{h}}\left(1-\frac{1}{2} \mu^{2} r_{h}^{2}\right),\\
A_{0}(r)=\mu-\mu \frac{r^{2}}{r_{h}^{2}}.
\end{gathered}
\end{equation}
The quark-antiquark pair is connected by U-shape string in the confined phase. If we choose the static gauge $\xi^{1}=t$ and $\xi^{2}=r$, then the boundary conditions for $x(r)$ become
\begin{equation}
x^{(1)}(0)=-\frac{L}{2}, \quad x^{(2)}(0)=\frac{L}{2}, x^{(i)}=0.
\end{equation}
By using the static gauge $t=\tau, \sigma=x$, the Nambu-Goto action of the U-shape string is
\begin{equation}\label{sng}
S=\frac{1}{2 \pi \alpha^{\prime}} \int d \tau d \sigma \sqrt{\operatorname{det} g_{\alpha \beta}},
\end{equation}
with
\begin{equation}
g_{\alpha \beta}=G_{\mu v} \frac{\partial x^{\mu}}{\partial \sigma^{\alpha}} \frac{\partial x^{v}}{\partial \sigma^{\beta}},
\end{equation}
where $X^{\mu}$ and $G_{\mu v}$ are the target space coordinates and the metric respectively, and $\sigma^{\alpha}$ with ¦Á = 0, 1 parameterize
the worldsheet.
Combining Eq.(\ref{sng}) and Eq.(\ref{metric0}), we can get
\begin{equation}
\begin{gathered}
S= \frac{g}{T} \int_{-\frac{L}{2}}^{\frac{L}{2}} d x \frac{e^{s r^{2}}}{r^{2}} \sqrt{f(r)+\left(\partial_{x} r\right)^{2}}.
\end{gathered}
\end{equation}
We choose the model parameters as follows:$g=\frac{R^{2}}{2 \pi a^{\prime}}$.
Now we set $w(r)=\frac{e^{s r^{2}}}{r^{2}}$ and identify the Lagrangian as
\begin{equation}
\mathcal{L}=w(r) \sqrt{f(r)+\left(\partial_{x} r\right)^{2}}
\end{equation}
and
\begin{equation}
\mathcal{L}-r^{\prime} \frac{\partial\mathcal{L}}{\partial r^{\prime}}=\frac{w(r) f(r)}{\sqrt{f(r)+\left(\partial_{x} r\right)^{2}}}=\rm constant.
\end{equation}
Using the conservation of energy, we have $\frac{w(r) f(r)}{\sqrt{f(r)+\left(\partial_{x} r\right)^{2}}}=w\left(r_{0}\right) \sqrt{f\left(r_{0}\right)}$ at the maximum point $r_{0}$ of the U-shape string. Thus, we get
\begin{equation}
\partial_{r} x=\sqrt{\frac{w^{2}\left(r_{0}\right) f^{2}\left(r_{0}\right)}{w^{2}(r) f^{2}(r) f\left(r_{0}\right)-w^{2}\left(r_{0}\right) f^{2}\left(r_{0}\right) f(r)}}.
\end{equation}
Finally, the separate length of quark-antiquark pair can be expressed as
\begin{equation}
\begin{gathered}
L=2 \int_{0}^{r_{0}} \partial_{r} x d r\\
=2 \int_{0}^{r_{0}} \sqrt{\frac{w^{2}\left(r_{0}\right) f^{2}\left(r_{0}\right)}{w^{2}(r) f^{2}(r) f\left(r_{0}\right)-w^{2}\left(r_{0}\right) f^{2}\left(r_{0}\right) f(r)}} d r.
\end{gathered}
\end{equation}
Considering $E = T S$, the regularized potential energy is
\begin{equation}
\begin{gathered}
E = \\
2 g \int_{0}^{r_{0}} (w(r) \sqrt{1+f(r) \frac{w^{2}\left(r_{0}\right) f^{2}\left(r_{0}\right)}{w^{2}(r) f^{2}(r) f\left(r_{0}\right)-w^{2}\left(r_{0}\right) f^{2}\left(r_{0}\right) f(r)}} \\
-\frac{1}{r^{2}} )d r-2 \frac{g}{r_{0}}.
\end{gathered}
\end{equation}

In our model, there is a dynamic wall in the confined phase. The string of quark-antiquark pair can not go beyond the wall and the system is in the confined phase. In the deconfined phase, the dynamic wall will disappear and the quark-antiquark pair can melt. Thus, we discuss the small black hole/large black hole phases which correspond to the specious-confined phase/deconfinement phases in our model. Then, the metric is valid in these different phases. More discussions can be found in \cite{Dudal:2017max,Li:2017tdz}.
\subsection{three-quark potential(A)}
The total action of the three-quark potential is the sum of the three Nambu-Goto actions plus the action of a vertex and boundary action $S_A|_{r=0}$
\begin{equation}
S=\sum_{i=1}^{3} S_{N G}^{(i)}+S_{\mathrm{vert}}+3 S_{A}.
\end{equation}
The baryon vertex is given by\cite{Chen:2021bkc,Andreev:2020pqy}
\begin{equation}
S_{\mathrm{vert}}=\tau_{v} \int \mathrm{d} t \frac{\mathrm{e}^{-2 s r^{2}}}{r} \sqrt{f(r)},
\end{equation}

where $\tau_{v}$ is a dimensionless parameter defined by $\tau_{v}=$ $\mathcal{T}_{5} R \operatorname{vol}(\boldsymbol{X})$, and $\operatorname{vol}(\boldsymbol{X})$ is a volume of $\boldsymbol{X}$. We redefine $\mathrm{k}=\frac{\tau_{v}}{3 g}$.

In the presence of a background gauge field, the string endpoints with attached quarks couple to it. So the world-sheet action includes boundary terms that are given by
\begin{equation}
S_{A}=\mp \frac{1}{3} \int d t A_{0}.
\end{equation}
The minus and plus signs correspond to a quark and an antiquark, respectively.

Thus, the action can now be written as
\begin{equation}
\begin{gathered}
S=\frac{3 g}{T} \int d x \frac{e^{s r^{2}}}{r^{2}} \sqrt{f(r)+\left(\partial_{x} r\right)^{2}}+\frac{3 \mathrm{k} g}{T} \frac{e^{-2 s r_{v}^{2}}}{r_{v}} \sqrt{f(r)}\\
-\frac{A_{0}(0)}{T}.
\end{gathered}
\end{equation}
The Lagrangian as
\begin{equation}
\mathcal{L}=\frac{e^{s r^{2}}}{r^{2}} \sqrt{f(r)+\left(\partial_{x} r\right)^{2}},
\end{equation}
and
\begin{equation}
\mathcal{L}-r^{\prime} \frac{\partial\mathcal{L}}{\partial r^{\prime}}=\frac{\frac{e^{s r^{2}}}{r^{2}} f(r)}{\sqrt{f(r)+\left(\partial_{x} r\right)^{2}}}=\rm constant.
\end{equation}
At the points $r_0$ and $r_v$, we have
\begin{equation}\label{20}
\begin{aligned}
&\frac{\frac{e^{s r^{2}}}{r^{2}} f(r)}{\sqrt{f(r)+\left(\partial_{x} r\right)^{2}}}=\frac{e^{s r_{0}^{2}}}{r_{0}^{2}} \sqrt{f\left(r_{0}\right)} \\
&\frac{\frac{e^{s r_{v}^{2}}}{r_{v}^{2}} f\left(r_{v}\right)}{\sqrt{f\left(r_{v}\right)+\tan ^{2} \alpha}}=\frac{e^{s r_{0}^{2}}}{r_{0}^{2}} \sqrt{f\left(r_{0}\right)}.
\end{aligned}
\end{equation}
The relation of $r_{0}, r_{v}$ and $\alpha$ can be determined from above equations. We can also get
\begin{equation}
\partial_{r} x=\sqrt{\frac{\frac{e^{2 s r_{0}^{2}}}{r_{0}^{4}} f^{2}\left(r_{0}\right)}{\frac{e^{2 s r^{2}}}{r^{4}} f^{2}(r) f\left(r_{0}\right)-\frac{e^{2 s r_{0}^{2}}}{r_{0}^{4}} f^{2}\left(r_{0}\right) f(r)}}
\end{equation}
and
\begin{equation}\label{22}
\frac{e^{s r_{v}^{2}}}{r_{v}^{2}} f\left(r_{v}\right)-\frac{e^{s r_{0}^{2}}}{r_{0}^{2}} \sqrt{f\left(r_{0}\right)} \sqrt{f\left(r_{v}\right)+\tan ^{2} \alpha}=0.
\end{equation}
The force balance equation at the point $r = r_v$ is
\begin{equation}
\bf{e_{1}+e_{2}+e_{3}+f_{v}}=0.
\end{equation}
Here $\bf{e_i}$ is the string tension, and $\bf{f_v}$ is a gravitational force acting on the vertex. The presence of this force is the main difference between string models in flat spaces and those in curved spaces. In this model, $\bf{f_v}$ only has one non-zero component in the r-direction. A formula for this component can be derived from the action $E_{\text {vert }}=T \cdot S_{\text {vert }}$. Explicitly, $f_{v}^{r}=-\delta E_{\text {vert }} / \delta r$ and $r$ is the coordinate of the vertex.
Each force is given by
\begin{equation}
\begin{gathered}
\boldsymbol{f}_{v}=\left(0,0,-3 g \mathrm{k} \partial_{r_{v}} \frac{e^{-2 s r_{v}^{2}}}{r_{v}} \sqrt{f\left(r_{v}\right)}\right) \\
\boldsymbol{e}_{\mathbf{1}}=-g \frac{e^{s r_{v}^{2}}}{r_{v}^{2}}\Bigg(\cos \beta \frac{f\left(r_{v}\right)}{\sqrt{f\left(r_{v}\right)+\tan ^{2} \alpha}},\\
\sin \beta \frac{f\left(r_{v}\right)}{\sqrt{f\left(r_{v}\right)+\tan ^{2} \alpha}},-\frac{1}{\sqrt{1+f\left(r_{v}\right) \cot ^{2} \alpha}}\Bigg) \\
\boldsymbol{e}_{2}=-g \frac{e^{s r_{v}^{2}}}{r_{v}^{2}}\Bigg(-\cos \beta \frac{f\left(r_{v}\right)}{\sqrt{f\left(r_{v}\right)+\tan ^{2} \alpha}}, \\
\sin \beta \frac{f\left(r_{v}\right)}{\sqrt{f\left(r_{v}\right)+\tan ^{2} \alpha}},-\frac{1}{\sqrt{1+f\left(r_{v}\right) \cot ^{2} \alpha}}\Bigg) \\
\boldsymbol{e}_{3}=-g \frac{e^{s r_{v}^{2}}}{r_{v}^{2}}\Bigg(0,-\frac{f\left(r_{v}\right)}{\sqrt{f\left(r_{v}\right)+\tan ^{2} \alpha}},\\-\frac{1}{\sqrt{1+f\left(r_{v}\right) \cot ^{2} \alpha}}\Bigg) \\
\end{gathered}
\end{equation}
Considering the symmetry in $x$ and $y$, we only need to consider the force balance in the r direction. The force balance equation leads to
\begin{equation}
\frac{e^{s r_{v}^{2}}}{r_{v}^{2}} \frac{1}{\sqrt{1+f\left(r_{v}\right) \cot ^{2} \alpha}}-\mathrm{k} \partial_{r_{v}} \frac{e^{-2 s r_{v}^{2}}}{r_{v}} \sqrt{f\left(r_{v}\right)}=0.
\end{equation}
Therefore, $L$ and $E$ can be calculated as
\begin{equation}
\begin{gathered}
L=\sqrt{3}\left(\int_{0}^{r_{0}} \partial_{r} x d r+\int_{r_{v}}^{r_{0}} \partial_{r} x d r\right)\\
=\sqrt{3}\int_{0}^{r_{0}} \sqrt{\frac{\frac{e^{2 s r_{0}^{2}}}{r_{0}^{4}} f^{2}\left(r_{0}\right)}{\frac{e^{2 s r^{2}}}{r^{4}} f^{2}(r) f\left(r_{0}\right)-\frac{e^{2 s r_{0}^{2}}}{r_{0}^{4}} f^{2}\left(r_{0}\right) f(r)}} d r \\
+\sqrt{3}\int_{r_{v}}^{r_{0}} \sqrt{\frac{\frac{e^{2 s r_{0}^{2}}}{r_{0}^{4}} f^{2}\left(r_{0}\right)}{\frac{e^{2 s r^{2}}}{r^{4}} f^{2}(r) f\left(r_{0}\right)-\frac{e^{2 s r_{0}^{2}}}{r_{0}^{4}} f^{2}\left(r_{0}\right) f(r)}} d r,
\end{gathered}
\end{equation}
\begin{equation}
\begin{gathered}
E=3 g(\int_{0}^{r_{0}} \frac{e^{s r^{2}}}{r^{2}} \sqrt{1+f(r)\left(\partial_{r} x\right)^{2}}-\frac{1}{r^{2}} d r+\\
\int_{r_{v}}^{r_{0}} \frac{e^{s r^{2}}}{r^{2}} \sqrt{1+f(r)\left(\partial_{r} x\right)^{2}} d r)\\
-\frac{3 g}{r_{0}}+3 g \mathrm{k} \frac{e^{-2 s r_{v}^{2}}}{r_{v}} \sqrt{f\left(r_{v}\right)}+3 c-\mu.
\end{gathered}
\end{equation}

\subsection{three-quark potential(B)}
We continue to discuss a connected collinear configuration of triply heavy baryon. Following the procedures in the last section, we can obtain the force balance equation
\begin{equation}
-2 \frac{e^{s r_{v}^{2}}}{r_{v}^{2}} \frac{1}{\sqrt{1+f\left(r_{v}\right) \cot ^{2} \alpha}}+\frac{e^{s r_{v}^{2}}}{r_{v}^{2}}+3k \partial_{r_{ v}} \frac{e^{-2 s r_{v}^{2}}}{r_{v}} \sqrt{f\left(r_{v}\right)}=0.
\end{equation}
Together with Eq.~(\ref{22}), the relation of $r_{0}$, $r_{v}$ and $\alpha$ can be determined. $\partial_{r} x$ in this configuration can be found in Eq.~(\ref{20}). Thus, we can similarly get inter-quark distance
\begin{equation}
\begin{aligned}
L &=\int_{0}^{r_{0}} \partial_{r} x d r+\int_{r_{v}}^{r_{0}} \partial_{r} x d r \\
&=\int_{0}^{r_{0}} \sqrt{\frac{\frac{e^{2 s r_{0}^{2}}}{r_{0}^{4}} f^{2}\left(r_{0}\right)}{\frac{e^{2 s r^{2}}}{r^{4}} f^{2}(r) f\left(r_{0}\right)-\frac{e^{2 s r_{0}^{2}}}{r_{0}^{4}} f^{2}\left(r_{0}\right) f(r)}} d r\\
&+\int_{r_{v}}^{r_{0}} \sqrt{\frac{\frac{e^{2 s r_{0}^{2}}}{r_{0}^{4}} f^{2}\left(r_{0}\right)}{\frac{e^{2 s r^{2}}}{r^{4}} f^{2}(r) f\left(r_{0}\right)-\frac{e^{2 s r_{0}^{2}}}{r_{0}^{4}} f^{2}\left(r_{0}\right) f(r)}} d r,
\end{aligned}
\end{equation}
and the potential energy
\begin{equation}
\begin{aligned}
\begin{gathered}
E= \\
2 g\left(\int_{0}^{r_{0}} \frac{e^{s r^{2}}}{r^{2}} \sqrt{1+f(r)\left(\partial_{r} x\right)^{2}} d r+\int_{r_{v}}^{r_{0}} \frac{e^{s r^{2}}}{r^{2}} \sqrt{1+f(r)\left(\partial_{r} x\right)^{2}} d r\right) \\
-\frac{2 g}{r_{0}}+g \int_{0}^{r_{v}} \frac{e^{s r^{2}}}{r^{2}}-\frac{1}{r^{2}} d r-\frac{g}{r_{v}}+3 g k \frac{e^{-2 s r_{v}^{2}}}{r_{v}} \sqrt{f(r_{v})}+3 c-\mu.
\end{gathered}
\end{aligned}
\end{equation}

\section{Numerical results }\label{Numerical results}
In this section, we will present the results based on the discussion above. First, we use this model to calculate the potential energy of the triply heavy baryon and quark-antiquark pair in Fig.\ref{2}. The parameters are fixed by the two-quark potential and three-quark potential of lattice results at vanishing temperature and chemical potential.
\begin{center}
	\includegraphics[width=8cm]{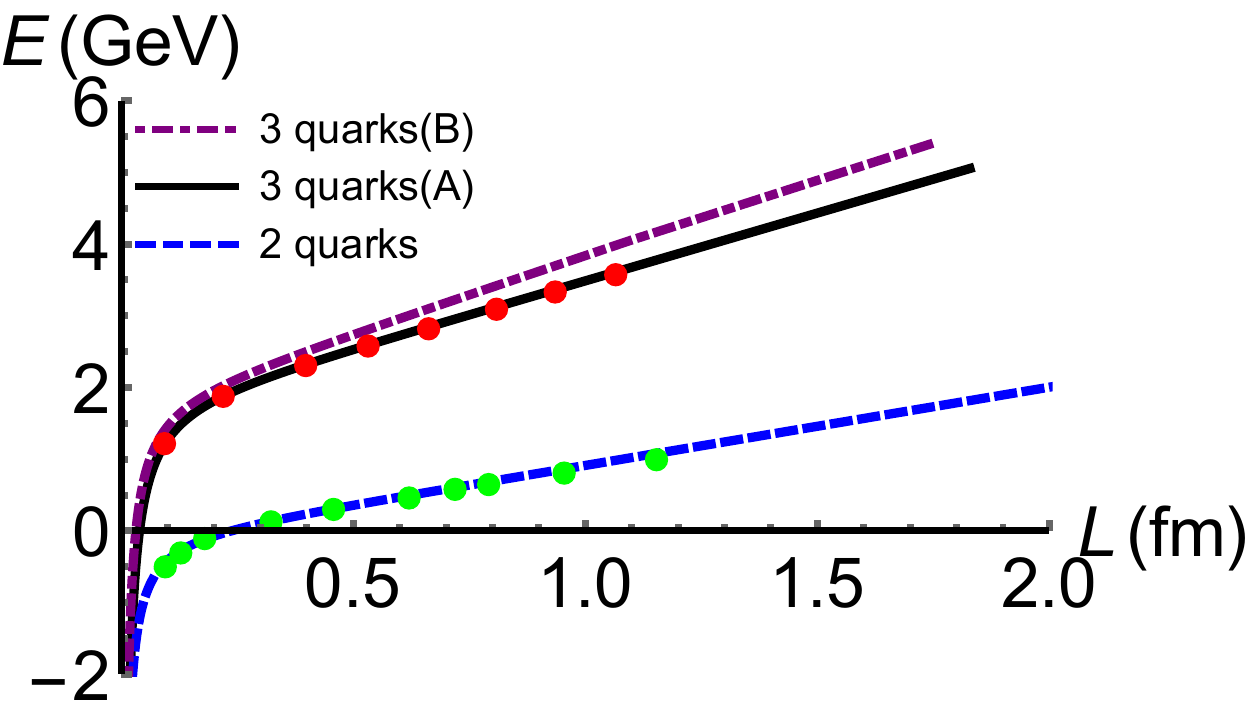}
	\figcaption{\label{2} The potential of quark-antiquark pair and triply heavy baryons as a function of separate distance at vanishing temperature. The dots are the results of lattice QCD\cite{Alexandrou:2002sn,Kaczmarek:2005ui}. Here s=0.450GeV,$g=0.176$,k=-0.102 and c=0.623GeV.}
\end{center}
Next, we investigate the behavior of separate distance at finite temperature and chemical potential in Fig.~\ref{3Lt} and Fig.~\ref{3Lut012}. The behavior of three-quark separate distance of configuration A is similar to quark-antiquark potential qualitatively. With the increase of temperature and chemical potential, the triply heavy baryon will melt. The screening distance can be read from the maximum of the dashed line.
\begin{center}
	\includegraphics[width=8cm]{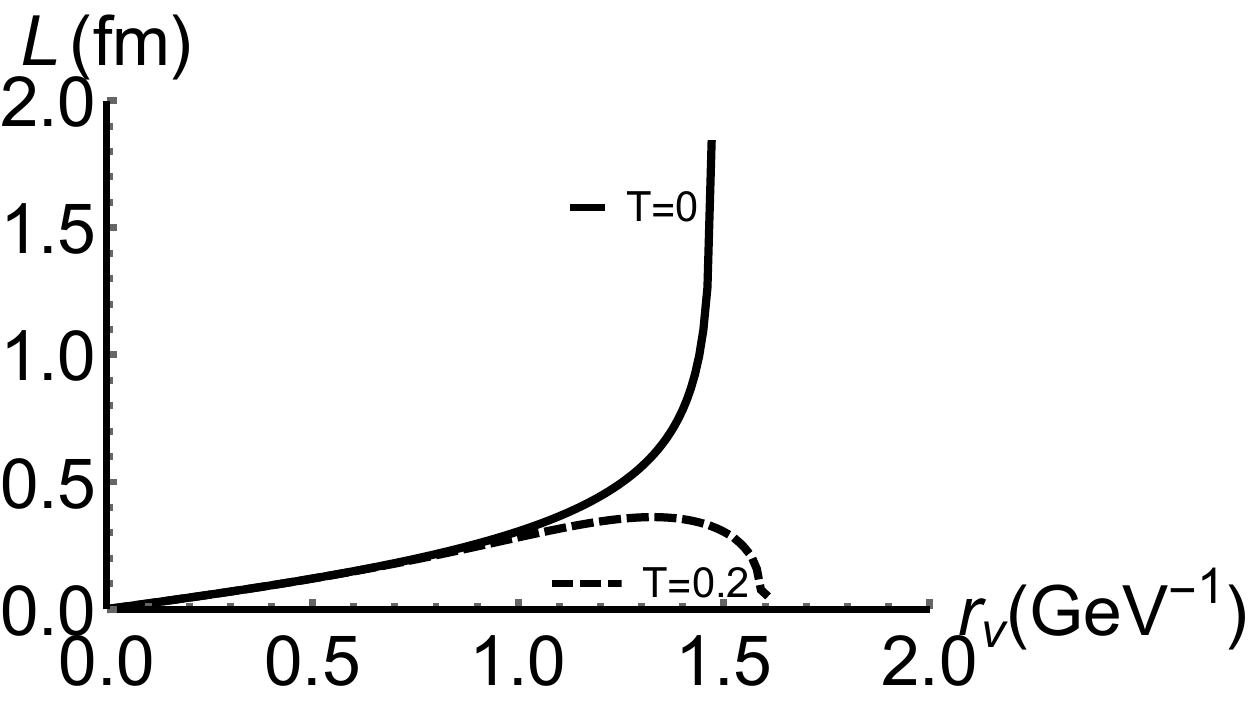}
    \figcaption{\label{3Lt} Separate distance of triply heavy baryon as a function of $r_v$ at vanishing chemical potential. The unit of temperature is GeV. }
\end{center}

Then, the two-quark and three-quark potential(A) are investigated at different temperatures and chemical potentials as shown in Fig.~\ref{23u0tbi}. With the increase of temperature and/or chemical potential, the triply heavy baryon and quark-antiquark pair dissolve. The melting temperature of the triply heavy baryon is 0.136 GeV which is very close to the melting temperature(0.138 GeV) of quark-antiquark pair. The screening distance $r_{d}$ of triply heavy baryon is 1.14 fm at the temperature 0.14GeV and $r_{d}$ is 0.45 fm at the temperature 0.18GeV in Fig.~\ref{23u0tbi}(a). The screening distance $r_{d}$ of quark-antiquark pair is 1.3 fm at the temperature 0.14GeV and $r_{d}$ is 0.5 fm at the temperature 0.18GeV in Fig.~\ref{23u0tbi}(b). It is found that the screening distance of the quark-antiquark pair is larger than that of triply heavy baryon. Thus, we infer that the quark-antiquark pair may be more stable than triply heavy baryon. Similarly, we fix the temperature and study the effect of chemical potential on 3-quark potential in Fig.~\ref{3ET012u}. It is found that the triply heavy baryon will also melt at large enough$\mu$.
\begin{center}
	\includegraphics[width=8cm]{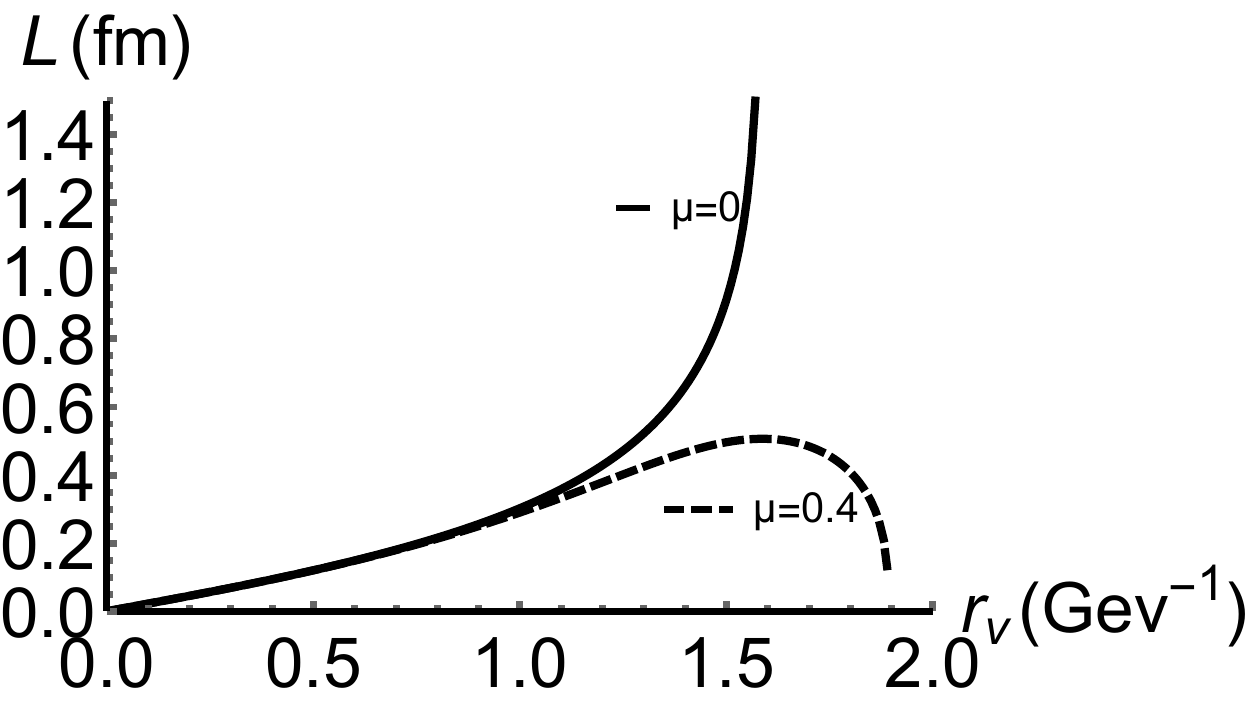}
    \figcaption{\label{3Lut012} Separate distance of triply heavy baryon as a function of $r_v$ at T=0.12GeV. The unit of $\mu$ is GeV.}
\end{center}
\end{multicols}
\begin{center}
    \includegraphics[width=16cm]{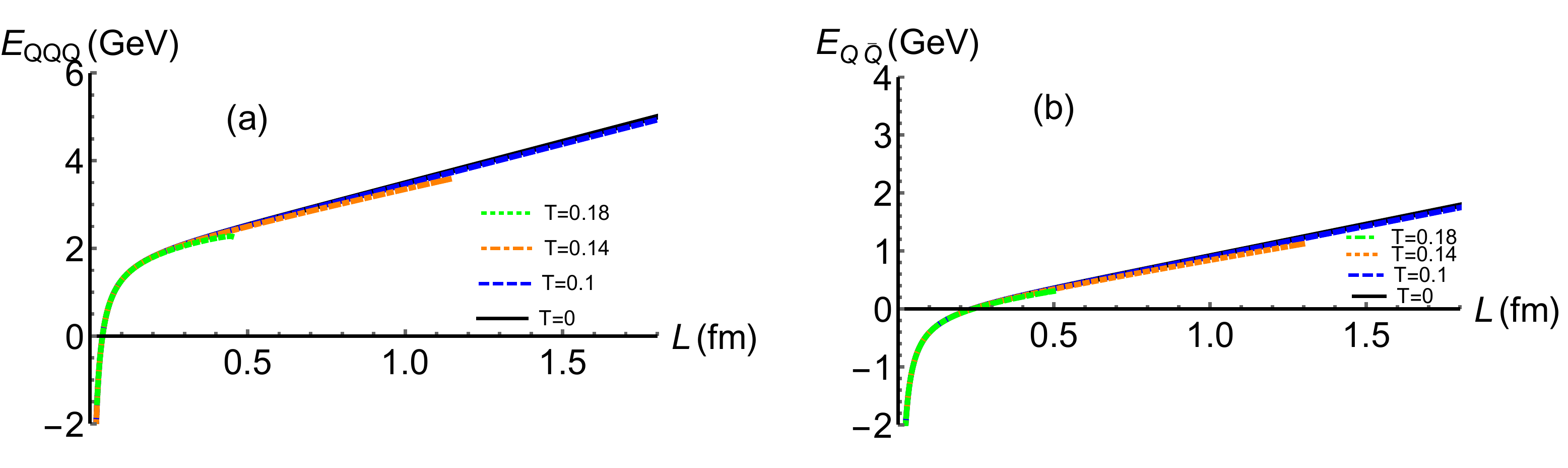}
    \figcaption{\label{23u0tbi}(a) Three-quark potential(A) as a function of separate distance L at different temperatures. (b) Quark-antiquark potential as a function of separate distance L at different temperatures. The chemical potential is vanishing and the unit of temperature is GeV.}
\end{center}
\begin{multicols}{2}
\begin{center}
   \includegraphics[width=8cm]{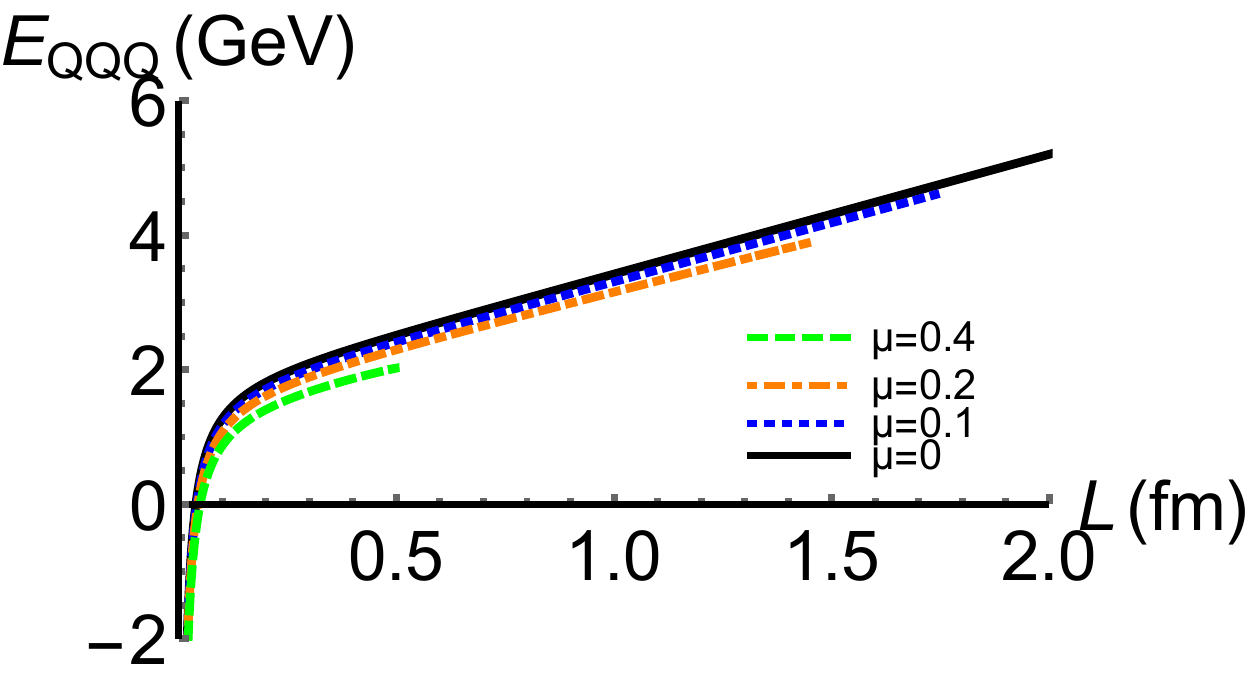}
   \figcaption{\label{3ET012u}Three-quark potential energy(A) as a function of separate distance L at different chemical potentials. The temperature is fixed as 0.12 GeV and the unit of chemical potential is GeV.}
\end{center}
To be more clear, we show the screening distance as a function of temperature/chemical potential in Fig.~\ref{23fl}. It is found that the screening distance becomes smaller with the increase of the temperature/chemical potential. Fig.~\ref{23fl}(a) shows that the screening distance decreases quickly at the temperature below 0.2GeV and then decreases slowly at high temperatures. Comparing with Fig.~\ref{23fl}, we conclude that the effect of temperature on screening distance is more significant than chemical potential.
\end{multicols}
\begin{center}
    \includegraphics[width=16cm]{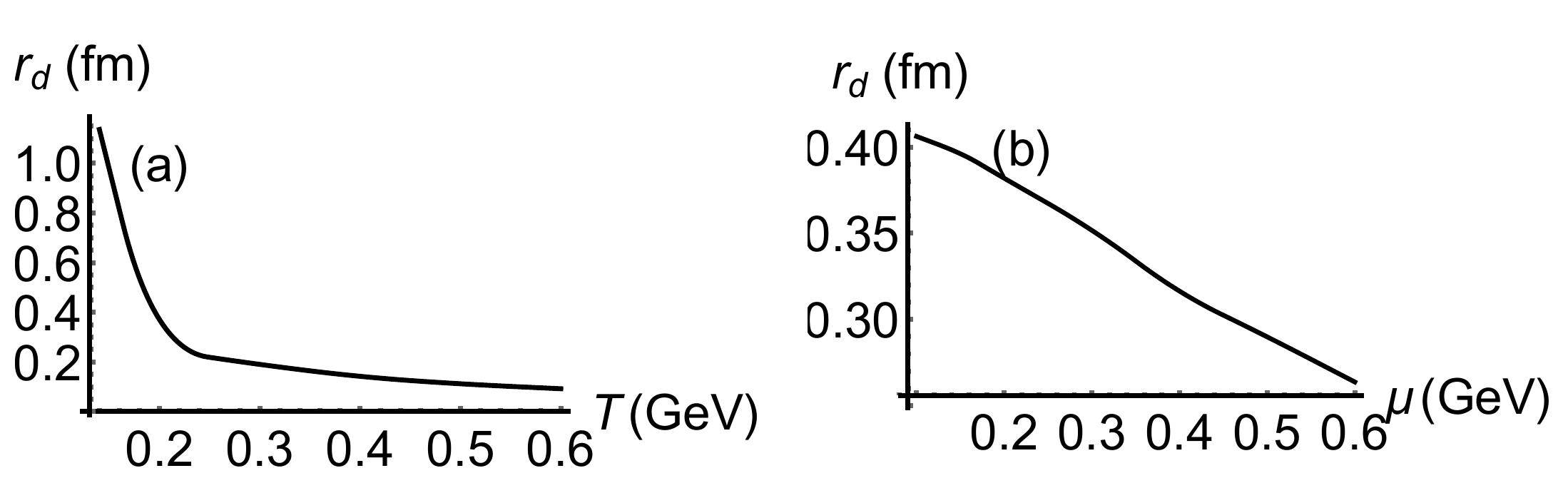}
    \figcaption{\label{23fl}(a) The screening distance $r_{d}$ as a function of temperature T at $\mu$=0. (b) The screening distance $r_{d}$ as a function of chemical potential $\mu$ at $T$=0.2GeV.}
\end{center}
\begin{multicols}{2}
\begin{center}
    \includegraphics[width=8cm]{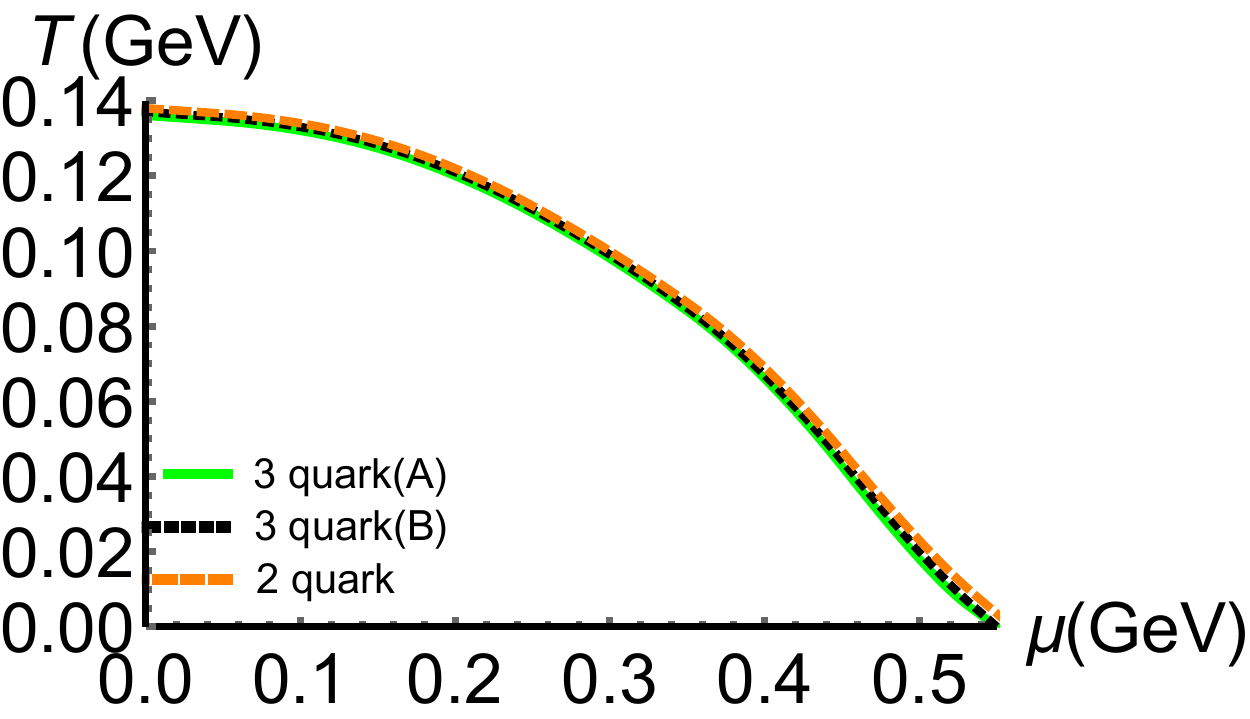}
    \figcaption{\label{3Tu}3-quark and 2-quark melting diagram in the $T-\mu$ plane.}
\end{center}
At last, the melting diagram of quark-antiquark pair and triply heavy baryon are shown in Fig.\ref{3Tu}. The melting temperature and chemical potential of the quark-antiquark pair are close to those of the triply heavy baryon. Even so, the three lines in Fig.\ref{3Tu} are not totally overlapped. Thus, we again infer that the quark-antiquark pair should be more stable than the triply heavy baryon.
\section{Summary and conclusion}\label{Summary and conclusion}
We study the potential energy of the triply heavy baryon at finite temperature and chemical potential in this paper. First, the calculation of three-quark potential is presented. With these in hand, numerical results have been done in the next part. At vanishing temperature, we fit the lattice results of potential energy and then calculate the separate distance of quarks at finite temperature and chemical potential. The potential energy of triply heavy baryon is shown in the confined and deconfined phases. From the figures of potential energy, we can determine the screening distance. It is found that the screening distance of the quark-antiquark pair is larger than that of triply heavy baryon at the same temperature and chemical potential. At last, the melting diagram of heavy quark-antiquark pair and triply heavy baryon are shown. Although three lines in the melting diagram are close, we can see the triply heavy baryon is easier to dissolve than the quark-antiquark pair. Moreover, Involving the light quarks is an interesting topic. We can continue to discuss the string breaking of QQQ at finite temperature and chemical potential. In the presence of light quark, we can investigate the string breaking of QQQ. The possible decay channels include QQQ $\rightarrow$ QQq+ Q$\rm \bar q$, QQQ $\rightarrow$ Qqq+ 2Q$\rm \bar q$ and QQQ $\rightarrow$ qqq + 3Q$\rm \bar q$. We will give detailed discussions for future work.
\section{Acknowledgments}
\acknowledgments{This work is supported by the NSFC under Grants No. 12175100, No. 11975132, the Research Foundation of Education Bureau of Hunan Province, China(Grant No. 21B0402 and No. 20C1594) and the Natural Science Foundation of Hunan Province of China under Grants No.2022JJ40344.}

\end{multicols}

\vspace{10mm}

\vspace{-1mm}
\centerline{\rule{80mm}{0.1pt}}

\end{CJK*}

\vspace{2mm}
\begin{thebibliography}{90}


\vspace{3mm}
\bibitem{Flynn:2011gf}
J.~M.~Flynn, E.~Hernandez and J.~Nieves,
Phys. Rev. D \textbf{85}, 014012 (2012)
doi:10.1103/PhysRevD.85.014012
[arXiv:1110.2962 [hep-ph]].

\bibitem{Andreev:2015riv}
O.~Andreev,
Phys. Rev. D \textbf{93}, no.10, 105014 (2016)
doi:10.1103/PhysRevD.93.105014
[arXiv:1511.03484 [hep-ph]].

\bibitem{Sommer:1984xq}
R.~Sommer and J.~Wosiek,
Phys. Lett. B \textbf{149}, 497-500 (1984)
doi:10.1016/0370-2693(84)90374-5

\bibitem{Sommer:1985da}
R.~Sommer and J.~Wosiek,
Nucl. Phys. B \textbf{267}, 531-538 (1986)
doi:10.1016/0550-3213(86)90129-X

\bibitem{Bali:2000gf}
G.~S.~Bali,
Phys. Rept. \textbf{343}, 1-136 (2001)
doi:10.1016/S0370-1573(00)00079-X
[arXiv:hep-ph/0001312 [hep-ph]].

\bibitem{Brambilla:2009cd}
N.~Brambilla, J.~Ghiglieri and A.~Vairo,
Phys. Rev. D \textbf{81}, 054031 (2010)
doi:10.1103/PhysRevD.81.054031
[arXiv:0911.3541 [hep-ph]].


\bibitem{Eichmann:2016yit}
G.~Eichmann, H.~Sanchis-Alepuz, R.~Williams, R.~Alkofer and C.~S.~Fischer,
Prog. Part. Nucl. Phys. \textbf{91}, 1-100 (2016)
doi:10.1016/j.ppnp.2016.07.001
[arXiv:1606.09602 [hep-ph]].

\bibitem{Borisenko:2018zzd}
O.~Borisenko, V.~Chelnokov, E.~Mendicelli and A.~Papa,
Nucl. Phys. B \textbf{940}, 214-238 (2019)
doi:10.1016/j.nuclphysb.2019.02.002
[arXiv:1812.05384 [hep-lat]].

\bibitem{Richard:2012xw}
J.~M.~Richard,
[arXiv:1205.4326 [hep-ph]].

\bibitem{deForcrand:2005vv}
P.~de Forcrand and O.~Jahn,
Nucl. Phys. A \textbf{755}, 475-480 (2005)
doi:10.1016/j.nuclphysa.2005.03.127
[arXiv:hep-ph/0502039 [hep-ph]].


\bibitem{Bornyakov:2004uv}
V.~G.~Bornyakov \textit{et al.} [DIK],
Phys. Rev. D \textbf{70}, 054506 (2004)
doi:10.1103/PhysRevD.70.054506
[arXiv:hep-lat/0401026 [hep-lat]].

\bibitem{Sakumichi:2015rfa}
N.~Sakumichi and H.~Suganuma,
Phys. Rev. D \textbf{92}, no.3, 034511 (2015)
doi:10.1103/PhysRevD.92.034511
[arXiv:1501.07596 [hep-lat]].

\bibitem{Bernard:2000gd}
C.~W.~Bernard, T.~Burch, K.~Orginos, D.~Toussaint, T.~A.~DeGrand, C.~E.~DeTar, S.~A.~Gottlieb, U.~M.~Heller, J.~E.~Hetrick and B.~Sugar,
Phys. Rev. D \textbf{62}, 034503 (2000)
doi:10.1103/PhysRevD.62.034503
[arXiv:hep-lat/0002028 [hep-lat]].

\bibitem{Takahashi:2000te}
T.~T.~Takahashi, H.~Matsufuru, Y.~Nemoto and H.~Suganuma,
Phys. Rev. Lett. \textbf{86}, 18-21 (2001)
doi:10.1103/PhysRevLett.86.18
[arXiv:hep-lat/0006005 [hep-lat]].

\bibitem{Matsufuru:2000qm}
H.~Matsufuru, Y.~Nemoto, H.~Suganuma, T.~T.~Takahashi and T.~Umeda,
Nucl. Phys. B Proc. Suppl. \textbf{94}, 554-557 (2001)
doi:10.1016/S0920-5632(01)00865-9
[arXiv:hep-lat/0010071 [hep-lat]].

\bibitem{Takahashi:2001nw}
T.~T.~Takahashi, H.~Suganuma, H.~Matsufuru and Y.~Nemoto,
AIP Conf. Proc. \textbf{594}, no.1, 341-348 (2002)
doi:10.1063/1.1425520
[arXiv:hep-lat/0107008 [hep-lat]].

\bibitem{Alexandrou:2001ip}
C.~Alexandrou, P.~De Forcrand and A.~Tsapalis,
Phys. Rev. D \textbf{65}, 054503 (2002)
doi:10.1103/PhysRevD.65.054503
[arXiv:hep-lat/0107006 [hep-lat]].

\bibitem{Takahashi:2002bw}
T.~T.~Takahashi, H.~Suganuma, Y.~Nemoto and H.~Matsufuru,
Phys. Rev. D \textbf{65}, 114509 (2002)
doi:10.1103/PhysRevD.65.114509
[arXiv:hep-lat/0204011 [hep-lat]].

\bibitem{Takahashi:2003ty}
T.~T.~Takahashi, H.~Matsufuru, Y.~Nemoto and H.~Suganuma,
[arXiv:hep-lat/0304009 [hep-lat]].

\bibitem{Takahashi:2004rw}
T.~T.~Takahashi and H.~Suganuma,
Phys. Rev. D \textbf{70}, 074506 (2004)
doi:10.1103/PhysRevD.70.074506
[arXiv:hep-lat/0409105 [hep-lat]].

\bibitem{Suganuma:2015xby}
H.~Suganuma and N.~Sakumichi,
PoS \textbf{LATTICE2015}, 323 (2016)
doi:10.22323/1.251.0323
[arXiv:1511.05244 [hep-lat]].


\bibitem{Alexandrou:2002sn}
C.~Alexandrou, P.~de Forcrand and O.~Jahn,
Nucl. Phys. B Proc. Suppl. \textbf{119}, 667-669 (2003)
doi:10.1016/S0920-5632(03)01659-1
[arXiv:hep-lat/0209062 [hep-lat]].


\bibitem{Koma:2017hcm}
Y.~Koma and M.~Koma,
Phys. Rev. D \textbf{95}, no.9, 094513 (2017)
doi:10.1103/PhysRevD.95.094513
[arXiv:1703.06247 [hep-lat]].





\bibitem{Maldacena:1997re}
J.~M.~Maldacena,
Adv. Theor. Math. Phys. \textbf{2}, 231-252 (1998)
doi:10.1023/A:1026654312961
[arXiv:hep-th/9711200 [hep-th]].

\bibitem{Gubser:1998bc}
S.~S.~Gubser, I.~R.~Klebanov and A.~M.~Polyakov,
Phys. Lett. B \textbf{428}, 105-114 (1998)
doi:10.1016/S0370-2693(98)00377-3
[arXiv:hep-th/9802109 [hep-th]].

\bibitem{Witten:1998qj}
E.~Witten,
Adv. Theor. Math. Phys. \textbf{2}, 253-291 (1998)
doi:10.4310/ATMP.1998.v2.n2.a2
[arXiv:hep-th/9802150 [hep-th]].


\bibitem{Maldacena:1998im}
J.~M.~Maldacena,
Phys. Rev. Lett. \textbf{80}, 4859-4862 (1998)
doi:10.1103/PhysRevLett.80.4859
[arXiv:hep-th/9803002 [hep-th]].


\bibitem{Rey:1998bq}
S.~J.~Rey, S.~Theisen and J.~T.~Yee,
Nucl. Phys. B \textbf{527}, 171-186 (1998)
doi:10.1016/S0550-3213(98)00471-4
[arXiv:hep-th/9803135 [hep-th]].


\bibitem{Andreev:2006nw}
O.~Andreev and V.~I.~Zakharov,
JHEP \textbf{04}, 100 (2007)
doi:10.1088/1126-6708/2007/04/100
[arXiv:hep-ph/0611304 [hep-ph]].


\bibitem{He:2010ye}
S.~He, M.~Huang and Q.~S.~Yan,
Phys. Rev. D \textbf{83}, 045034 (2011)
doi:10.1103/PhysRevD.83.045034
[arXiv:1004.1880 [hep-ph]].

\bibitem{Colangelo:2010pe}
P.~Colangelo, F.~Giannuzzi and S.~Nicotri,
Phys. Rev. D \textbf{83}, 035015 (2011)
doi:10.1103/PhysRevD.83.035015
[arXiv:1008.3116 [hep-ph]].

\bibitem{Li:2011hp}
D.~Li, S.~He, M.~Huang and Q.~S.~Yan,
JHEP \textbf{09}, 041 (2011)
doi:10.1007/JHEP09(2011)041
[arXiv:1103.5389 [hep-th]].

\bibitem{Fadafan:2012qy}
K.~B.~Fadafan and E.~Azimfard,
Nucl. Phys. B \textbf{863}, 347-360 (2012)
doi:10.1016/j.nuclphysb.2012.05.022
[arXiv:1203.3942 [hep-th]].

\bibitem{Chakraborty:2012dt}
S.~Chakraborty and N.~Haque,
Nucl. Phys. B \textbf{874}, 821-851 (2013)
doi:10.1016/j.nuclphysb.2013.06.010
[arXiv:1212.2769 [hep-th]].


\bibitem{Finazzo:2014rca}
S.~I.~Finazzo and J.~Noronha,
JHEP \textbf{01}, 051 (2015)
doi:10.1007/JHEP01(2015)051
[arXiv:1406.2683 [hep-th]].

\bibitem{Zhang:2015faa}
Z.~q.~Zhang, D.~f.~Hou and G.~Chen,
Nucl. Phys. A \textbf{960}, 1-10 (2017)
doi:10.1016/j.nuclphysa.2017.01.007
[arXiv:1507.07263 [hep-ph]].

\bibitem{Gursoy:2020kjd}
U.~G\"ursoy, M.~J\"arvinen, G.~Nijs and J.~F.~Pedraza,
JHEP \textbf{03}, 180 (2021)
doi:10.1007/JHEP03(2021)180
[arXiv:2011.09474 [hep-th]].


\bibitem{Chen:2017lsf}
X.~Chen, S.~Q.~Feng, Y.~F.~Shi and Y.~Zhong,
Phys. Rev. D \textbf{97}, no.6, 066015 (2018)
doi:10.1103/PhysRevD.97.066015
[arXiv:1710.00465 [hep-ph]].

\bibitem{Zhou:2020ssi}
J.~Zhou, X.~Chen, Y.~Q.~Zhao and J.~Ping,
Phys. Rev. D \textbf{102}, no.8, 086020 (2020)
doi:10.1103/PhysRevD.102.086020
[arXiv:2006.09062 [hep-ph]].

\bibitem{Zhou:2021sdy}
J.~Zhou, X.~Chen, Y.~Q.~Zhao and J.~Ping,
Phys. Rev. D \textbf{102}, no.12, 126029 (2021)
doi:10.1103/PhysRevD.102.126029


\bibitem{Andreev:2015iaa}
O.~Andreev,
Phys. Lett. B \textbf{756}, 6-9 (2016)
doi:10.1016/j.physletb.2016.02.070
[arXiv:1505.01067 [hep-ph]].

\bibitem{Andreev:2019cbc}
O.~Andreev,
Phys. Lett. B \textbf{804}, 135406 (2020)
doi:10.1016/j.physletb.2020.135406
[arXiv:1909.12771 [hep-ph]].

\bibitem{Andreev:2020xor}
O.~Andreev,
JHEP \textbf{05}, 173 (2021)
doi:10.1007/JHEP05(2021)173
[arXiv:2007.15466 [hep-ph]].

\bibitem{Andreev:2021bfg}
O.~Andreev,
Phys. Rev. D \textbf{104}, no.2, 026005 (2021)
doi:10.1103/PhysRevD.104.026005
[arXiv:2101.03858 [hep-ph]].

\bibitem{Andreev:2021eyj}
O.~Andreev,
Phys. Rev. D \textbf{105}, no.8, 086025 (2022)
doi:10.1103/PhysRevD.105.086025
[arXiv:2111.14418 [hep-ph]].


\bibitem{Andreev:2022cax}
O.~Andreev,
[arXiv:2205.12119 [hep-ph]].

\bibitem{Chen:2021bkc}
X.~Chen, B.~Yu, P.~C.~Chu and X.~h.~Li,
Chin. Phys. C \textbf{46}, no.7, 073102 (2022)
doi:10.1088/1674-1137/ac5db9
[arXiv:2112.06234 [hep-ph]].

\bibitem{Li:2017tdz}
M.~W.~Li, Y.~Yang and P.~H.~Yuan,
Phys. Rev. D \textbf{96}, no.6, 066013 (2017)
doi:10.1103/PhysRevD.96.066013
[arXiv:1703.09184 [hep-th]].

\bibitem{Dudal:2017max}
D.~Dudal and S.~Mahapatra,
Phys. Rev. D \textbf{96}, no.12, 126010 (2017)
doi:10.1103/PhysRevD.96.126010
[arXiv:1708.06995 [hep-th]].


\bibitem{Andreev:2020pqy}
O.~Andreev,
Phys. Rev. D \textbf{101}, no.10, 106003 (2020)
doi:10.1103/PhysRevD.101.106003
[arXiv:2003.09880 [hep-ph]].

\bibitem{Kaczmarek:2005ui}
O.~Kaczmarek and F.~Zantow,
Phys. Rev. D \textbf{71}, 114510 (2005)
doi:10.1103/PhysRevD.71.114510
[arXiv:hep-lat/0503017 [hep-lat]].



\end{thebibliography}
\end{document}